\documentclass[fleqn,twoside]{article}
\usepackage{espcrc2}
\usepackage{amsmath}
\usepackage{amssymb}
\usepackage{amsfonts}
\usepackage{citesort}

\usepackage{graphicx}
\usepackage[figuresright]{rotating}

\title{Results of NEMO~3 and Status of SuperNEMO}

\author{L.~V\'ala\address[CTU]{Czech Technical University in Prague,
    IEAP, Horsk\'a 3a/22, CZ -- 128 00 Prague, Czech Republic}%
  \thanks{E-mail: {\tt ladislav.vala@utef.cvut.cz}}
  on behalf of the NEMO and SuperNEMO Collaborations}
               
\begin{document}

\begin{abstract}
  The NEMO~3 experiment is devoted 
  to the search for neutrinoless double beta decay,
  as well as for accurate measurement of two-neutrino double beta decay.
  The detector has been taking data in the LSM laboratory since 2003
  and the latest NEMO~3 results for seven $\beta\beta$ isotopes
  are presented here for both decay modes.
  The SuperNEMO project aims to extend the NEMO technique
  to a 100 -- 200~kg isotope experiment 
  with the target half-life sensitivity of $1 - 2 \times 10^{26}$~y.
  The current status of the SuperNEMO R\&D programme is described.
\end{abstract}

\maketitle

\section{Introduction}
The currently running NEMO~3 experiment \cite{ARN05a}
(NEMO = Neutrino Ettore Majorana Observatory) 
is devoted to the search for 
neutrinoless double beta decay ($0\nu\beta\beta$) and 
to the accurate measurement of 
two-neutrino double beta decay ($2\nu\beta\beta$)
by means of the direct detection of the two electrons.
$0\nu\beta\beta$ decay is a process beyond Standard Model 
because of the violation of lepton number conservation by two units.
Detection of this process is the best experimental test 
for Majorana character of neutrinos.
This kind of decay constrains also the type of mass spectrum hierarchy
and the absolute mass of the neutrinos.

\section{NEMO~3 detector}
The NEMO~3 detector is located  
in the Modane underground laboratory (LSM) in the Fr\'ejus tunnel
in France at the depth of 4800~m w.e.
The set-up is cylindrical in design,
it is divided into twenty equal sectors
and combines two detection techniques --
particle identification provided by a wire tracking chamber
and energy and time measurements of particles 
with a calorimeter.
Thus, NEMO~3 is able to identify
$e^-$, $e^+$, photons, and $\alpha$-particles
which allows recognition of different $\beta\beta$ decay modes,
measurement of internal and external backgrounds,
as well as good discrimination between signal and background.

The tracking detector is made of 6180 open octagonal drift cells
operated in Geiger mode 
and providing a three-dimensional measurement of 
the charged particle tracks 
by recording the drift time and the two plasma propagation times.
The calorimeter, which surrounds the wire chamber,
consists of 1940 plastic scintillators coupled 
to very low-radioactivity PMTs
and gives an energy resolution of $14 - 17\%/\sqrt{E}$ FWHM at 1~MeV
and the time resolution of 250~ps.

Seventeen sectors of NEMO~3 accommodate almost 10~kg of 
enriched $\beta\beta$ isotopes (see Table~\ref{tab:results-2nubb}) 
in the form of thin foils. 
Three sectors are also used for
external background measurement
and are equipped with pure Cu and natural Te.

The detector is surrounded by a solenoidal coil generating 
magnetic field of 25~Gauss
for the $e^- / e^+$ recognition,
and is covered by two types of shielding 
against external $\gamma$-rays and neutrons.

\section{Event selection and background}
A candidate for a $\beta\beta$ decay is a two-electron event
which is selected by requiring
two reconstructed tracks with 
a curvature corresponding to the negative charge and 
coming from the same vertex in the source foils.
Each track has to be associated with a fired scintillator,
energy of each $e^-$ 
measured in the calorimeter should be higher than 200~keV
and the time-of-flight has to correspond to 
the case of two electrons emitted at the same time 
from the common vertex in the foils.

A complete study of background 
in the $0\nu\beta\beta$ window has been performed. 
The level of each background component has been directly measured
from data using different analysis channels.
The background can be classified in three groups:
1) internal radioactive contamination of the source,
2) external background from incoming $\gamma$-rays and 
3) radon inside the tracking volume.

The dominant background during the first running period
from February 2003 to September 2004 (Phase~I)
was due to radon diffusion into the wire chamber 
through tiny air leaks.
The radon level inside NEMO~3 during the second running period
after installation of the radon trapping facility in November 2004
(Phase~II) has been reduced by a factor of ten.
Remaining low radon activity inside NEMO~3 
is due to detector component degassing.

\section{NEMO~3 results}
Measurements of the $2\nu\beta\beta$ decay half-lives 
have been performed for 
all the seven $\beta\beta$ isotopes in NEMO~3.
The obtained half-lives, 
combining new preliminary results for 
$^{150}$Nd, $^{130}$Te, $^{96}$Zr, and $^{48}$Ca 
with higher statistics (Phase I and II data)
and previous results for $^{100}$Mo and $^{82}$Se (Phase I data),
are given in Table~\ref{tab:results-2nubb}.

No evidence was found for the $0\nu\beta\beta$ decay of
$^{100}$Mo, $^{82}$Se, $^{150}$Nd, $^{96}$Zr, and $^{48}$Ca.
The $0\nu\beta\beta$ decay half-life limits and 
limits on the effective Majorana neutrino mass $\langle m_\nu \rangle$
have been derived and are summarised in Table~\ref{tab:results-0nubb}.

\begin{table*}
  \caption{$\beta\beta$ isotopes installed in NEMO~3 and results for $2\nu\beta\beta$ decay half-life measurement.}
  \label{tab:results-2nubb}
  \begin{tabular}{cr@{}lclr@{}l}
    \hline
    Isotope & \multicolumn{2}{l}{Mass (g)}& Q$_{\beta\beta}$ (keV) & $T_{1/2}^{2\nu}$ (y) & \multicolumn{2}{l}{$S/B$}\\
    \hline
    $^{100}$Mo & 6914 & & 3034 & $[7.11 \pm 0.02(stat) \pm 0.54(syst)] \times 10^{18}$ \cite{ARN05b} & 40 & \\
    $^{82}$Se  &  932 & & 2995 & $[9.6 \pm 0.3(stat) \pm 1.0(syst)] \times 10^{19}$ \cite{ARN05b} & 4 & .0 \\
    $^{150}$Nd &  37&.0 & 3367 & $[9.11^{+0.25}_{-0.22}(stat) \pm 0.63(syst)] \times 10^{18}$ \cite{ARG08} & 2 & .8 \\   
    $^{130}$Te &  454 & & 2529 & $[7.6 \pm 1.5(stat) \pm 0.8(syst)] \times 10^{20} $    & 0 & .25 \\
    $^{116}$Cd &  405 & & 2805 & $[2.8 \pm 0.1(stat) \pm 0.3(syst)] \times 10^{19} $    & 7 & .5 \\
    $^{96}$Zr  &  9 &.4 & 3350 & $[2.3 \pm 0.2(stat) \pm 0.3(syst)] \times 10^{19} $    & 1 & .0 \\
    $^{48}$Ca  &  7 &.0 & 4272 & $[4.4^{+0.5}_{-0.4}(stat) \pm 0.4(syst)] \times 10^{19} $    & 6& .8  \\
    \hline
  \end{tabular} 
\end{table*}

\begin{table*}
    \caption{NEMO~3 results: $0\nu\beta\beta$ decay half-life limits at 90\% C.L.}
    \label{tab:results-0nubb}
    \begin{tabular}{cclll}
      \hline
      Isotope & $0\nu\beta\beta$ mode & $T_{1/2}^{0\nu}$ limit (90\% C.L.) &  & NME\\
      \hline
      $^{100}$Mo & (V -- A) 
                 & $> 5.8 \times 10^{23}$~y \cite{ARN05b}
                 & $\langle m_{\nu} \rangle < 0.8 - 1.3$~eV & \cite{KOR07a-KOR07b,ROD07} \\
                 & (V + A) 
                 & $> 3.2 \times 10^{23}$~y \cite{ARN05b} & & \\
      $^{82}$Se  & (V -- A) 
                 & $> 2.1 \times 10^{23}$~y \cite{ARN05b}
                 & $\langle m_{\nu} \rangle < 1.4 - 2.2$~eV &  \cite{KOR07a-KOR07b,ROD07} \\
                 & (V + A) 
                 & $> 1.2 \times 10^{23}$~y \cite{ARN05b} & & \\
      $^{150}$Nd & (V -- A) 
                 & $> 1.80 \times 10^{22}$~y \cite{ARG08}
                 & $\langle m_{\nu} \rangle < 3.7 - 5.1$~eV & \cite{ROD06} \\
                 & (V + A) 
                 & $> 1.07 \times 10^{22}$~y \cite{ARG08} & & \\
      $^{96}$Zr  & (V -- A) 
                 & $> 8.6 \times 10^{22}$~y 
                 & $\langle m_{\nu} \rangle < 7.4 - 20.1$~eV & \cite{KOR07a-KOR07b,SIM08} \\
      $^{48}$Ca  & (V -- A) 
                 & $> 1.3 \times 10^{22}$~y 
                 & $\langle m_{\nu} \rangle < 29.6$~eV & \cite{CAU08} \\
      \hline
    \end{tabular} 
 \end{table*} 

\section{SuperNEMO detector design}
SuperNEMO aims to extend and improve the experimental techniques
use by the current NEMO~3 experiment
in order to search for $0\nu\beta\beta$ decay
with a target half-life sensitivity of 
$1-2 \times 10^{26}$~year,
which corresponds to the effective neutrino mass sensitivity 
$\langle m_{\nu} \rangle$ of $40 - 100$~meV, 
depending on nuclear matrix elements (NME) used.
The SuperNEMO project is a $\sim 100-200$~kg source experiment
and is currently in a three year design study and R\&D phase. 

Like NEMO~3, SuperNEMO will combine calorimetry and tracking.
This technological choice allows the measurement of
individual electron tracks, event vertices, energies and time-of-flight,
and thus provides the full reconstruction of kinematics and topology of an event.
SuperNEMO will consist of about twenty identical modules,
each housing around 5 -- 7~kg of isotope
in the form of thin foil ($\sim$ 40~mg/cm$^2$).
The tracking volume contains more than 2000 wire drift cells 
operated in Geiger mode (Geiger cells),
which are arranged in nine layers parallel to the foil.
The calorimeter is divided into $\sim$ 1000 blocks,
which cover most of the detector outer area
and are read out by low background PMTs.

\section{SuperNEMO R\&D}
The R\&D programme focuses on four main areas of study:
(i) calorimeter, (ii) isotope enrichment, (iii) tracking detector, and
(iv) ultra-low background materials and measurements.

SuperNEMO aims to improve the calorimeter energy resolution to
$7\%/\sqrt{E}$ FWHM at 1~MeV.
To reach this goal,
several ongoing studies are investigating 
the choice of calorimeter parameters such as
scintillator material (plastic or liquid)
and the shape, size and coating of calorimeter blocks \cite{KAU08}.
These are combined with dedicated development of PMTs 
with very low radioactivity and high quantum efficiency.
The collaboration expects 
to make the final decision on the calorimeter design in mid-2009.

The tracking detector design study has for its goal
optimisation of the wire chamber parameters
to obtain high efficiency and resolution in measuring 
the trajectories of electrons from $\beta\beta$ decay, 
as well as of $\alpha$-particles for the purpose of background rejection.
The first 9-cell prototype was successfully operated 
demonstrating propagation efficiency close to 100\%
over a wide range of voltages \cite{NAS07}.
Recently, a 90-cell prototype has been set in operation.

The main candidate $\beta\beta$ isotopes for SuperNEMO
are $^{82}$Se and $^{150}$Nd.
A sample of 4~kg of $^{82}$Se has been enriched 
and is currently undergoing purification.
The SuperNEMO collaboration is investigating
the possibility of enriching large amounts of $^{150}$Nd
via the atomic vapour laser isotope separation (AVLIS) method.

In order to reach required sensitivity,
SuperNEMO has to maintain ultra-low levels of background 
(one order lower than for NEMO~3).
Contamination of sources has to be less than 2~$\mu$Bq/kg for $^{208}$Tl
and less than 10~$\mu$Bq/kg for $^{214}$Bi.
In order to evaluate these activities,
a dedicated BiPo detector is being developed.
The first prototype, BiPo1,
installed in the LSM laboratory in February 2008,
reached the background level of 
$<7.5$~$\mu$Bq/kg for $^{208}$Tl (90\% C.L.) \cite{BON08}.
Another prototype, BiPo2, was installed in the LSM in July 2008.

\section{Conclusion}
The NEMO~3 detector has been routinely taking data since 2003.
The $2\nu\beta\beta$ decay half-lives of seven isotopes
have been measured with high statistics
and with better precision than in previous measurements.
No evidence for the $0\nu\beta\beta$ decay has been found in data
and the $T_{1/2}$ and $\langle m_{\nu} \rangle$ limits have been set.
The next generation experiment SuperNEMO will extrapolate 
the NEMO technique of calorimetry plus tracking
to 100 -- 200~kg of $\beta\beta$ isotope scale experiment.
Due to its modular design,
SuperNEMO can start operation in stages,
with the first module installed in 2011
and all the modules running by 2014.

%

%

\end{document}